\documentclass[12pt]{iopart}

\usepackage{graphicx}
\usepackage{bm} 
\usepackage{subfigure}

\begin{document}

\title[Nucleon-nucleon collision profile and ...]{
Nucleon-nucleon collision profile and cross section fluctuations}

\author{Maciej Rybczy\'nski}
\address{Institute of Physics, Jan Kochanowski University, PL-25406~Kielce, Poland}
\ead{Maciej.Rybczynski@ujk.edu.pl}

\author{Zbigniew W\l odarczyk}
\address{Institute of Physics, Jan Kochanowski University, PL-25406~Kielce, Poland}
\ead{Zbigniew.Wlodarczyk@ujk.edu.pl}

\begin{abstract}
The nucleon-nucleon collision profile, being the basic entity of the wounded nucleon model, is usually adopted in the form of hard sphere or the Gaussian shape. We suggest that the cross section fluctuations given by the gamma distribution leads to the profile function which smoothly ranges between the both limiting forms. Examples demonstrating sensitivity of profile function on cross section fluctuations are discussed. 
\end{abstract}

\vspace{2pc}

\submitto{\JPG}
\maketitle

\section{Introduction}
\label{sect:intro}

The wounded nucleon model~\cite{Bialas:1976ed} and its extensions~\cite{Kharzeev:2000ph, Broniowski:2007nz} have become the basic tool in an analysis of relativistic heavy ion collisions. This is essentially the bare framework of the traditional Glauber model~\cite{Glauber:1959aa}, with all the quantum mechanics reduced to its simplest form~\cite{Glauber:2006gd}. The original Glauber model develops into the purely classical, geometric picture used for present-day data analysis. The Glauber Monte Carlo calculations have been used for description of the bulk properties of the data~\cite{Back:2004dy, Bozek:2011wa, Miller:2007ri, Broniowski:2009fm}.

In the wounded nucleon model the basic entity is the nucleon-nucleon collision profile, $p\left(b\right)$, defined by the probability density $f\left(b\right)$ of inelastic nucleon-nucleon collision at the impact parameter $b$. It is normalized to the total inelastic nucleon-nucleon cross section:
\begin{equation}
2\pi \int db b p\left(b\right)=\sigma_{inel}
\label{eq1}
\end{equation}
In most Glauber Monte Carlo codes it is assumed that $p\left(b\right)$ is a simple step function
\begin{equation}
p\left(b\right)=\Theta\left(R-b\right),
\label{eq2}
\end{equation}
with $\sigma_{inel}=\pi\cdot R^2$ and probability density distribution $f\left(b\right)=\frac{1}{R}\Theta\left(R-b\right)$.

The importance of the shape of the nucleon-nucleon collision profile was discussed in Ref.~\cite{Rybczynski:2011wv}.
Following~\cite{Bialas:2006qf}, where the CERN ISR experimental data for pp cross section were properly parametrized with a combination of Gaussians, in Ref.~\cite{Rybczynski:2011wv} the profile function was proposed in the form
\begin{equation}
p\left(b\right)=A\exp\left(-\frac{\pi A b^2}{\sigma_{inel}}\right)
\label{eq4}
\end{equation}
with parameter $A=0.92$.

In this work we demonstrate the convenient way to model hadronic collision profile is by introducing the cross section fluctuations~\cite{Good:1960ba, Miettinen:1978jb, Kopeliovich:1981pz, Blaettel:1993ah, Frankfurt:1993qi, Strikman:1995jf, Guzey:2005tk, Heiselberg:1991is, Frankfurt:2008vi, Strikman:2011ar}. The main goal of the present work is to discuss a particular model of cross section fluctuations and its impact on measurements.

\section{Cross section fluctuations}
\label{sect:csf}

The Glauber formalism offers a convenient scheme for the calculation of various observables measured in the hadron-nucleus (or nucleus-nucleus) scattering at high energies such as the total and elastic cross sections
\begin{eqnarray}
\sigma^{hA}_{tot}& = & 2\int d^2 \vec{b}\: \Re\: G\left(\vec{b}\right)\nonumber,\\
\sigma^{hA}_{el} & = & \int d^2 \vec{b}\: \Re\: \Bigl\vert G\left(\vec{b}\right)\Bigl\vert^2,
\label{eq5}
\end{eqnarray}
where $G\left(\vec{b}\right)$ is the nuclear nuclear impact parameter elastic amplitude.

The origin of the cross section fluctuations can be traced down to the fact that hadrons have internal degrees of freedom (color-carrying quarks and gluons) and can therefore collide in different internal configurations resulting in different cross sections~\cite{Blaettel:1993ah, Frankfurt:2008vi}. Considering the incoming wave as a coherent superposition of eigenstates of the scattering operator, each eigenstate interacts with the target with its own cross section~\cite{Good:1960ba}. Since, in general, these cross sections (eigenvalues) are different, the final state contains not only the initial particle but also other states. It is important to note that the formalism of scattering eigenstates is based on the assumption that one can represent scattering as a superposition of the components with different interaction strengths. The use of this assumption and the completeness of the set of scattering states allows to obtain compact formulas. Introducing the probability of interaction with a given cross section $\sigma$, $f\left(\sigma\right)$, the expressions for the total and elastic hadron-nucleus cross sections become
\begin{eqnarray}
\sigma^{hA}_{tot} & = & 2\int f\left(\sigma\right)d\sigma\int d^2 \vec{b}\: \Re\: G\left(\vec{b},\sigma\right)\nonumber,\\
\sigma^{hA}_{el} & = & \int f\left(\sigma\right)d\sigma\int d^2 \vec{b}\: \Re\: \Bigl\vert G\left(\vec{b},\sigma\right)\Bigl\vert^2.
\label{eq6}
\end{eqnarray}
At small impact parameters and large $\sigma$, the nuclear scattering amplitude saturates, $G\left(\vec{b},\sigma\right)\cong 1$, and becomes independent of $\sigma$. Therefore cross section fluctuations indicate how close to the black body limit regime one is. In equations~(\ref{eq6}) the scattering amplitude $G\left(\vec{b},\sigma\right)$ depends on the eigenvalue $\sigma$ rather than on the total cross section $\sigma^{pp}_{tot}$, and the cross sections are sensitive not only to the first moment of $f\left(\sigma\right)$, $\langle\sigma\rangle=\sigma^{pp}_{tot}$ but also to the higher moments $\langle\sigma^{k}\rangle$.
The distribution over cross sections $f\left(\sigma\right)$ has the following properties:
\begin{eqnarray}
\int d\sigma f\left(\sigma\right) & = & 1,\nonumber\\
\int d\sigma \sigma f\left(\sigma\right) & = & \langle\sigma\rangle,\nonumber\\
\int d\sigma \sigma^2 f\left(\sigma\right) & = & \langle\sigma^2\rangle = \langle\sigma\rangle^2\left(1+\omega\right),
\label{eq8}
\end{eqnarray}
where
\begin{equation}
\omega=\frac{Var\left(\sigma\right)}{\langle\sigma\rangle^2}
\label{eq9}
\end{equation}
measures the broadness of cross section fluctuations around the average value.

As a first estimate a Gaussian distribution modified to give the correct asymptotic behavior for $\sigma\rightarrow 0$ was considered~\cite{Blaettel:1993ah,Guzey:2005tk}
\begin{equation}
f\left(\sigma\right) = N\left(\sigma_0,\Omega\right)\frac{\sigma}{\sigma + \sigma_0} \exp\left(-\frac{\left(\sigma-\sigma_0\right)^2}{\sigma^2_0\Omega^2}\right)
\label{eq10}
\end{equation}
where $\sigma_0$ and $\omega$ are the parameters, and $N\left(\sigma_0,\Omega\right)$ is the normalization constant.

It is well known~\cite{Wilk:1994mh, wilk_levy_flights} that cross section fluctuations lead to nonexponential attenuation of hadrons. As was shown~\cite{Wilk:1999dr,Wilk:2008ue} fluctuations given by the gamma distribution
\begin{equation}
g\left(\sigma\right)=\frac{1}{\Gamma\left(1/\omega\right)}\frac{1}{\sigma_0\omega}\left(\frac{1}{\omega}\frac{\sigma}{\sigma_0}\right)^{\frac{1}{\omega}-1}\exp\left(-\frac{1}{\omega}\frac{\sigma}{\sigma_0}\right)
\label{eq11}
\end{equation}
with $\Gamma\left(z\right)$ being Euler gamma function inevitably lead to the quasi power-law distribution of hadron free path
\begin{equation}
h\left(x\right)=\int_0^{\infty}d\sigma\frac{1}{\lambda\left(\sigma\right)}\exp \left(-\frac{x}{\lambda\left(\sigma\right)}\right)g\left(\sigma\right)=\frac{1-\omega}{\lambda_0\left(\sigma_0\right)}\left[1+\omega\frac{x}{\lambda_0\left(\sigma_0\right)}\right]^{1/\omega}
\label{eq12}
\end{equation}
characterized by the mean free path $\lambda_0\propto 1/\sigma_0$ and the parameter $\omega\in\left(0,1\right)$ reflecting the amount of fluctuations of cross section. Notice, that asymptotically (for $\omega\rightarrow 0$) the gamma distribution (\ref{eq11}) becomes delta function $\delta\left(\sigma-\sigma_0\right)$ and $h\left(x\right)\rightarrow \exp\left(-x/\lambda_{0}\right)$.

\section{Profile function}
\label{sect:pf}

Fluctuation of cross section, given by equation~(\ref{eq11}), results geometrically in fluctuations of nucleon radius. In the picture of hard-sphere approximation, the radius $R$ fluctuate as
\begin{equation}
f\left(R\right)=\frac{1}{\Gamma\left(1/\omega\right)}\frac{2}{R_{0}\sqrt{\omega}}\left(\frac{R}{R_{0}\sqrt{\omega}}\right)^{\frac{2}{\omega-1}}\exp\left(-\frac{R^2}{R_{0}^{2}\omega}\right).
\label{eq14}
\end{equation}
Equations~(\ref{eq1}) and~(\ref{eq8}) result in equalities:
\begin{equation}
\int_{0}^{\infty} d\sigma \sigma g\left(\sigma\right) = \int_{0}^{\infty} db 2\pi b p\left(b\right),
\label{eq15}
\end{equation}
\begin{equation}
\int_{0}^{\infty} d\sigma g\left(\sigma\right)\int_{0}^{\infty} db 2\pi b\Theta\left(\sigma-\pi b^{2}\right) = \int_{0}^{\infty} db 2\pi b p\left(b\right),
\label{eq16}
\end{equation}
\begin{equation}
\int_{0}^{\infty} db 2\pi b \left(\int_{\pi b^{2}}^{\infty}d\sigma g\left(\sigma\right)\right) = \int_{0}^{\infty} db 2\pi b p\left(b\right).
\label{eq17}
\end{equation}
Finally we can write
\begin{equation}
p\left(b\right) = \int_{\pi b^{2}}^{\infty} d\sigma g\left(\sigma\right).
\label{eq18}
\end{equation}
Integrating distribution~(\ref{eq11}) with $\sigma_{0}=\pi R_{0}^{2}$ we have
\begin{eqnarray}
p\left(b\right) & = & 1-\frac{1}{\Gamma\left(1/\omega\right)}\sum_{k=0}^{\infty}\frac{\left(-1\right)^{k}}{\left(k+\frac{1}{\omega}\right)\Gamma\left(k+1\right)}\left(\frac{b^2}{R_{0}^{2}\omega}\right)^{k+\frac{1}{\omega}}\nonumber\\ & = & \Gamma\left(\frac{1}{\omega},\frac{b^2}{R_{0}^{2}\omega}\right)/\Gamma\left(\frac{1}{\omega}\right),
\label{eq19}
\end{eqnarray}
where $\Gamma\left(\alpha,z\right)$ is the incomplete gamma function.

Profile function (\ref{eq19}) ranges from the hard sphere
\begin{equation*}
\lim_{\omega \to 0} p\left(b\right) = \Theta\left(R_{0}-b\right)
\end{equation*}
to the Gaussian shape
\begin{equation*}
\lim_{\omega \to 1} p\left(b\right) = \exp\left(-\frac{b^2}{R_{0}^{2}}\right)
\end{equation*}
Non-physical shapes for $\omega>1$ (large fluctuations in comparison with experimentally expected maximal values $\omega \propto 0.4$)~\cite{Blaettel:1993ah, Guzey:2005tk} are also possible in this scenario.

\begin{figure}[t]
\begin{center}
\includegraphics[angle=0,width=0.75\textwidth]{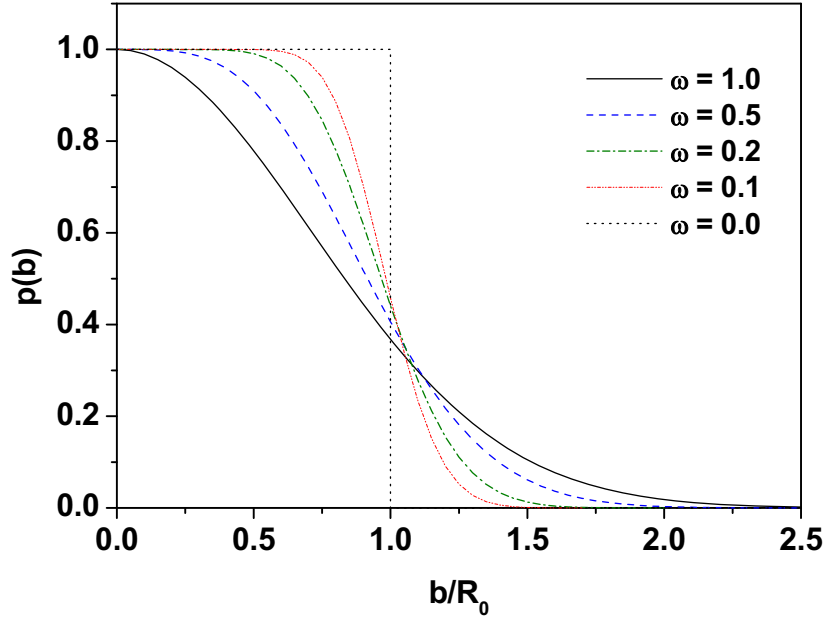}
\end{center}
\caption{\small (Color online) Nucleon-nucleon profile function $p\left(b\right)$ for different cross section fluctuations characterised by $\omega={\rm Var}\left(\sigma\right)/\langle\sigma\rangle^2$. See text for details.}
\label{fig1}
\end{figure}

Nucleon-nucleon profile functions~(\ref{eq19}) for different values of fluctuation parameter $\omega$ are shown in figure~\ref{fig1}. For $\omega=0$ we have profile for hard sphere (cf. equation~(\ref{eq2})) whereas for $\omega=1$ the profile is Gaussian (cf. equation~(\ref{eq4}) with $A=1$).

Recently~\cite{Brogueira:2010ep, Brogueira:2011jb}, the impact parameter elastic amplitude was proposed in the form
\begin{equation}
G\left(b\right)=\frac{1}{\exp\left(\frac{b-r_{0}}{\gamma}\right)+1}
\label{eq21}
\end{equation}
being solution of the logistic equation
\begin{equation}
\frac{\partial G\left(b\right)}{\partial b}=-\frac{1}{\gamma}\left(G\left(b\right)-G^{2}\left(b\right)\right),
\label{eq22}
\end{equation}
where $r_{0}$ and $\gamma$ are parameters.
It is interesting to notice that profile function $\alpha G\left(b\right)$ given by equation~(\ref{eq21}) with parameters 
\begin{eqnarray}
\alpha & \cong & 0.9855+0.11694\cdot\omega,\nonumber\\
r_{0}/R_{0} & \cong & 1.01387-0.23312\cdot\omega, \nonumber \\
\gamma/R_{0} & \cong & 0.07914+0.24546\cdot\omega \nonumber
\end{eqnarray}
is numerically consistent with $p\left(b\right)$ as given by equation~(\ref{eq19}).
\begin{figure}[t]
\begin{center}
\includegraphics[angle=0,width=0.75\textwidth]{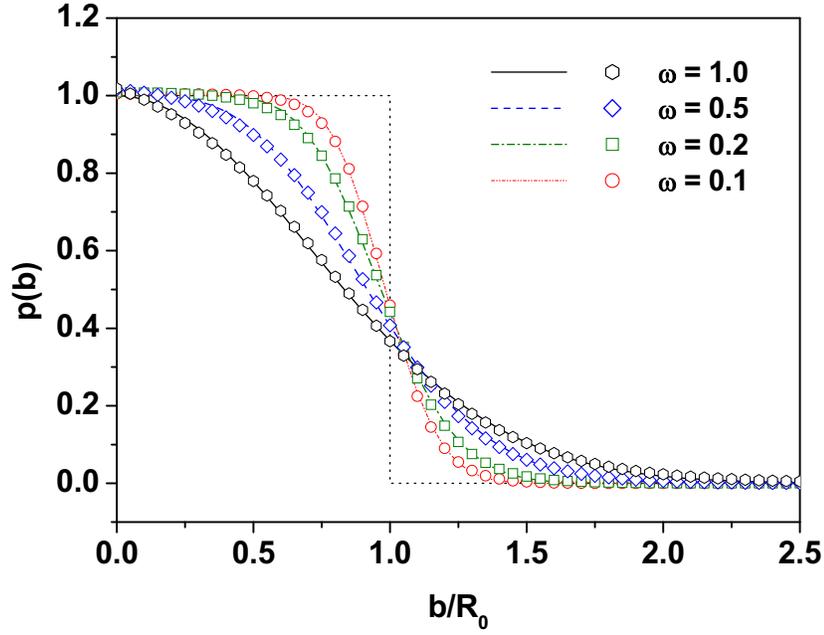}
\end{center}
\caption{\small (Color online) Nucleon-nucleon profile function $p\left(b\right)$ for different cross section fluctuations characterised by $\omega={\rm Var}\left(\sigma\right)/\langle\sigma\rangle^2$. Line comes from equation~(\ref{eq19}) and symbols from equation~(\ref{eq21}). See text for details.}
\label{fig2}
\end{figure}

\section{Information from elastic scattering}
\label{sect:ies}

The unitarity condition links the elastic nucleon-nucleon scattering amplitude, $t_{el}\left(b\right)$ with the inelastic collision profile\footnote{Here and in the following we ignore the real part of the amplitude.}
\begin{equation}
t_{el}\left(b\right)=1-\sqrt{1-p\left(b\right)}.
\label{eq23}
\end{equation}
The elastic amplitude in momentum transfer representation $T_{el}\left(q\right)$ is a Fourier
transform of the amplitude in impact parameter space:
\begin{equation}
T_{el}\left(q\right)=\int d^2b \exp\left(i\vec{b}\cdot\vec{q}\right)t_{el}\left(b\right).
\label{eq24}
\end{equation}
The total nucleon-nucleon cross section may be expressed as:
\begin{equation}
\sigma_{tot}=2T_{el}\left(0\right),
\label{eq25}
\end{equation}
and the total elastic cross section is:
\begin{equation}
\sigma_{el}=\sigma_{tot}-\sigma_{inel}=\int d^2b \vert t_{el}\left(b\right)\vert^{2}.
\label{eq26}
\end{equation}
Finally, the elastic differential cross section is given by:
\begin{equation}
\frac{d\sigma_{el}}{dt}=\frac{1}{4\pi} \Bigl\vert T_{el}\left(q\right)\Bigl\vert^{2},
\label{eq27}
\end{equation}
where $t\simeq -q^{2}$. 

\begin{figure}[t]
\begin{center}
\includegraphics[angle=0,width=0.99\textwidth]{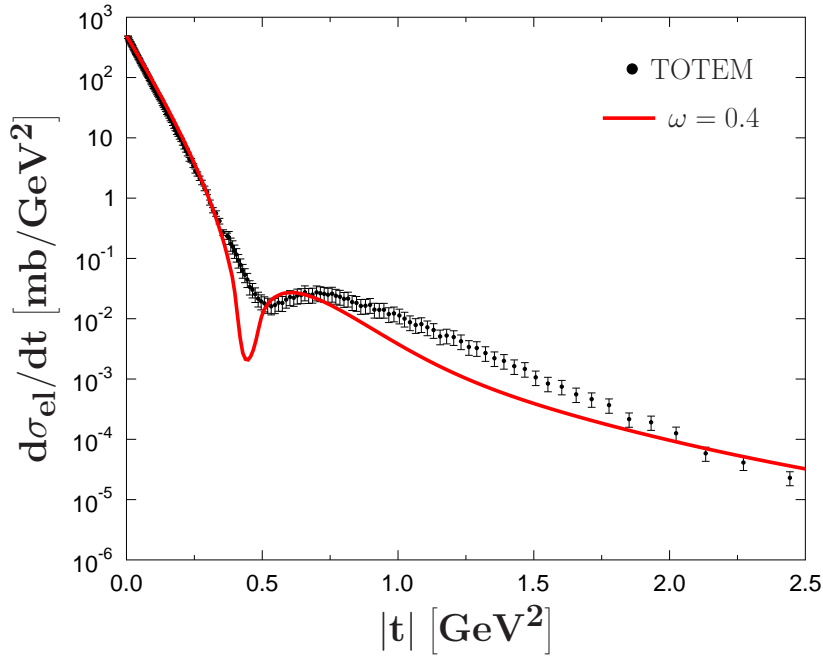}
\end{center}
\caption{\small (Color online) TOTEM experiment data on elastic differential proton-proton cross section~\cite{Antchev:2011zz, Antchev:2013gaa} (dots) fitted by equation~(\ref{eq27}) (line) with use of profile function $p\left(b\right)$ as given by equation~(\ref{eq19}). See text for details.}
\label{figmain}
\end{figure}

\begin{figure}[t]
\begin{center}
\includegraphics[angle=0,width=0.99\textwidth]{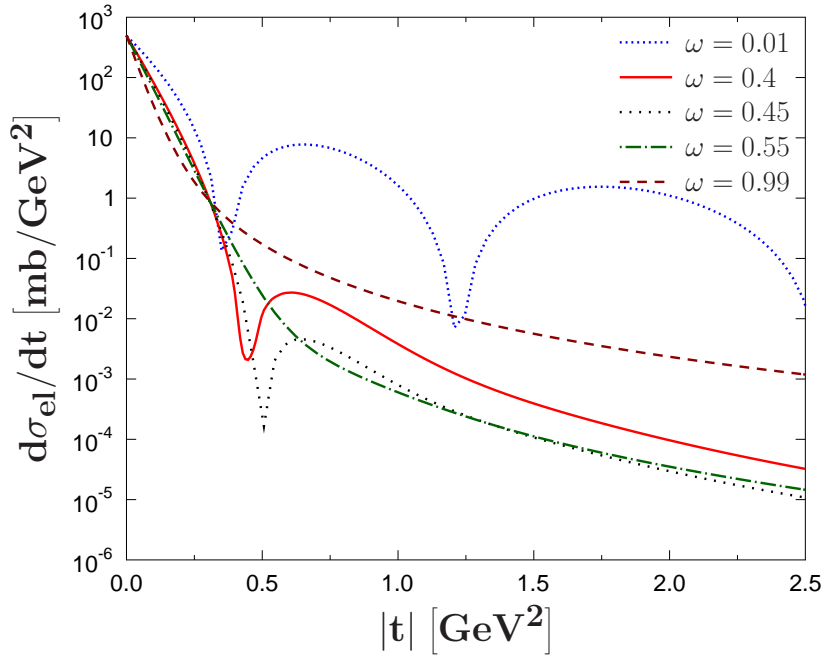}
\end{center}
\caption{\small (Color online) Set of solutions of equation~(\ref{eq27}) with use of profile function $p\left(b\right)$ as given by equation~(\ref{eq19}). All parameters besides $\omega$ are the same as in fugure~\ref{figmain}. See text for details.}
\label{figcurves}
\end{figure}

Recently, the TOTEM Collaboration published results on elastic differential proton-proton cross section at $\sqrt{s}=7$~TeV~\cite{Antchev:2011zz, Antchev:2013gaa}. The use of profile function $p\left(b\right)$ as given by equation~(\ref{eq19}) with the parameters: $\sigma_{inel}=73.2$~mb, $\omega=0.4$ and $R_{0}^{2}=k\cdot\sigma_{inel}/\pi$ ($k=0.97$ to recover the total proton-proton cross section, $\sigma_{tot}=98.3$ mb) allows for proper parametrization of TOTEM data, (see figure~\ref{figmain}).
Set of solutions of equation~(\ref{eq19}) for different $\omega$ is shown on the figure~\ref{figcurves}. 

From the above we can conclude that the profile function is determined by the different configuration of partons. The distribution of partons in momentum space, e.g. the number of wee partons in a hadron affects its cross section. The concept of cross section fluctuations takes into account all different configurations, regardless of their particular structure. Whether a configuration is "small" or "large" is not determined exclusively  by the spatial distribution of the partons. We can illustrate this taking into account the black disc with radius $R$ distributed according $f\left(R\right)$ given by equation~(\ref{eq14}). In this limit, $t_{el}\left(b\right)=1$ for $0 \leq b \leq R$, we have
\begin{equation}
T_{el}\left(q\right)=2\pi R^{2}\frac{J_{1}\left(R q\right)}{R q}
\label{eq28}
\end{equation}
and differential elastic cross section is given by
\begin{equation}
\frac{d\sigma_{el}}{dt}=\frac{1}{4\pi}\int_{0}^{\infty} dR \Bigl\vert T_{el}\left(q\right)\Bigl\vert^{2}  f\left(R\right)
\label{eq29}
\end{equation}
with $t\simeq -q^{2}$.

\begin{figure}[t]
\begin{center}
\includegraphics[angle=0,width=0.75\textwidth]{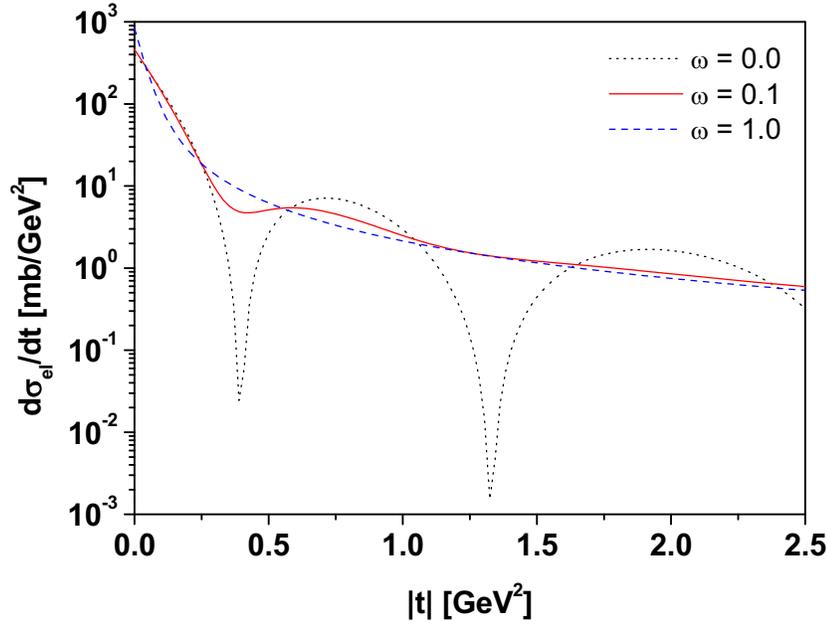}
\end{center}
\caption{\small (Color online) Set of solutions of equation~(\ref{eq29}) for different values of $\omega$. See text for details.}
\label{figblackdisk}
\end{figure}

Set of solutions of equation~(\ref{eq29}) for different $\omega$ is shown in figure~\ref{figblackdisk}.
Obtained cross sections (see figure~\ref{figblackdisk}) are remarkably higher than shown in figure~\ref{figcurves} and they cannot accommodate experimental data shown in figure~\ref{figmain}. Any fluctuations of cross section are not able to describe experimental data in this scenario.

\section{Implication for nuclear collisions}
\label{sect:inc}

\begin{figure}[t]
\begin{center}
\includegraphics[angle=0,width=0.99\textwidth]{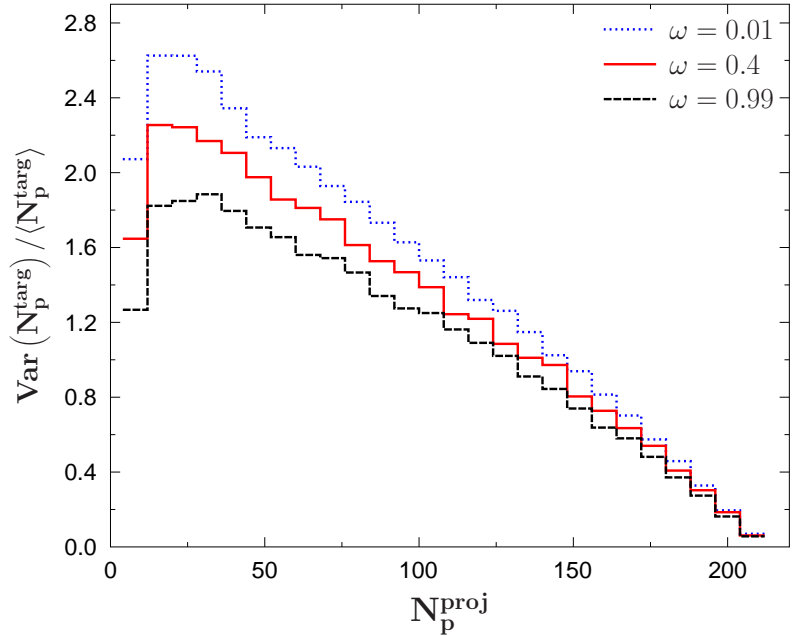}
\end{center}
\caption{\small (Color online) Scaled variance of number of target participants as a function of number of projectile participants for Pb+Pb minimum bias collisions at $\sigma_{inel}=73.2$~mb. See text for details.}
\label{figomegatarg}
\end{figure}

The use of different values of $\omega$ may affect various observables which can be measured experimentally. As example we demonstrate that cross section fluctuations lead to modification of the distribution of the number of nucleons in inelastic nucleus-nucleus collision. We show in figure~\ref{figomegatarg} a scaled variance of number of target participants as a function of number of projectile participants for Pb+Pb minimum bias collisions at $\sigma_{inel}=73.2$~mb for three values of $\omega$, almost hard sphere ($\omega=0.01$), almost Gaussian ($\omega=0.99$), and $\omega=0.4$ obtained from the fit of TOTEM data~\cite{Antchev:2011zz, Antchev:2013gaa}. There is about $50\%$ difference in magnitude of scaled variance between hard sphere and Gaussian approximation. The results presented in figure~\ref{figomegatarg} was prepared with use of Glauber Monte Carlo code {\tt GLISSANDO}~\cite{Broniowski:2007nz}.

Fluctuations in the number of target participants originate both from cross section fluctuations and from fluctuations of the number of nucleons along the path of projectile. In the framework of probabilistic approach to nucleus-nucleus collisions~\cite{Vechernin:2011mm} we can write for the mean value and for the variance  of $N_{p}^{targ}$ the following expressions:
\begin{equation}
\langle N_{p}^{targ}\rangle = A\left(1-Q\right) = N_{p}^{proj},
\label{eq:meannptarg}
\end{equation}
\begin{equation}
{\rm Var}\left(N_{p}^{targ}\right) = A Q\left(1-Q\right)+A\left(A-1\right)\left[\tilde{Q}-Q^{2}\right].
\label{eq:varnptarg}
\end{equation}
where $A$ is the atomic mass number and $Q=1-N_{p}^{proj}/A$ depends only on the nucleon-nucleon cross section $\langle\sigma\rangle$, but $\tilde{Q}$ depends also on the shape of the profile function $p\left(b\right)$. Roughly, $\tilde{Q}\left(\omega\right)\simeq Q^{2-\alpha\left(\omega\right)}$ with $\alpha\left(\omega\right)=10^{-3}\left(N_{p}^{proj}\right)^{1/4}\left(2\omega-4.46\right)$ and for increasing $\omega$ the second term in equation~(\ref{eq:varnptarg}) decrease.

It is remarkable that for increasing fluctuations of cross section (increasing $\omega$), the scaled fluctuations of $N_{p}^{targ}$ decrease.

\section{Multiplicty and cross section}
\label{sect:mcs}

\begin{figure}[t]
\begin{center}
\includegraphics[angle=0,width=0.75\textwidth]{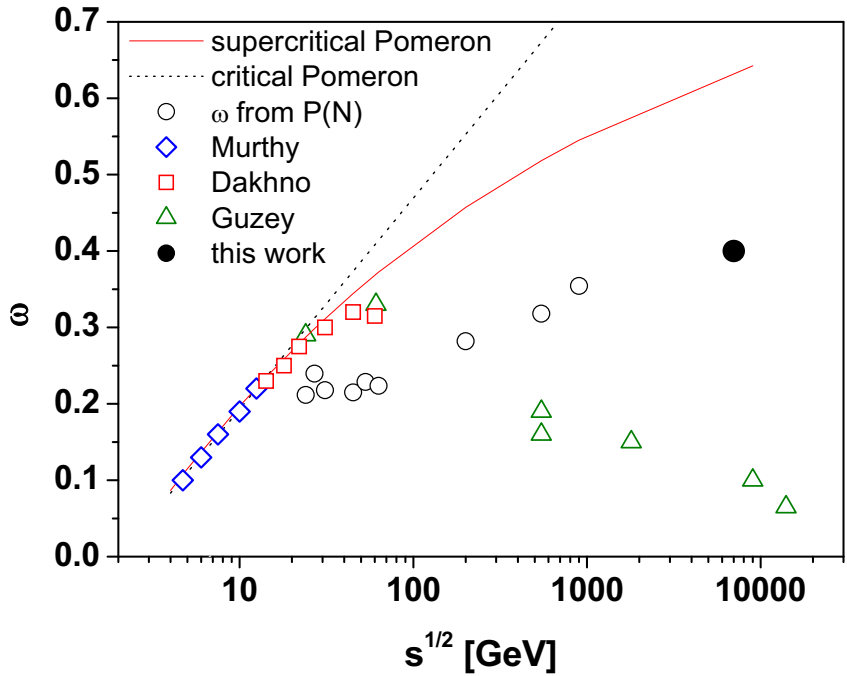}
\end{center}
\caption{\small (Color online) Nucleon-nucleon cross section fluctuations. The curves showing predictions based on critical single Pomeron exchange and the supercritical Pomeron are taken from~\cite{Blaettel:1993ah}. Data marked as Murthy and Dakeno comes from~\cite{Blaettel:1993ah} while data Guzey comes from~\cite{Guzey:2005tk}. Full circle show $\omega$ evaluated from elastic differential cross section and open circles show $\omega$ obtained from multiplicity fluctuations. See text for details.}
\label{fig3}
\end{figure}

Although constrains on the production cross section imposed by the observed bahavior of multiplicity distributions have been investigated already long time ago~\cite{Novero:1981bi, Yokomi:1976db, Yokomi:1976qi, Bialas:1977xd}, this subject is not pursued at moment. In Ref.~\cite{Rybczynski:2004gs} we demonstrate that $\langle N\rangle \propto \langle\sigma\rangle$. More accurate analysis~\cite{Yudong} indicate the relation
\begin{equation}
\langle N\rangle = C\left(s\right) \langle\sigma\rangle,
\label{eq30}
\end{equation}
where $C\left(s\right)=10^{-2}\left(-3.88+3.96\ln s\right)$.
Following~\cite{Yudong} we can assume that for given $\sigma$ we have Poissonian multiplicity distribution
\begin{equation}
P\left(N\right)=\frac{\bar{N}^{N}}{N!}\exp\left(-\bar{N}\right)
\label{eq31}
\end{equation}
and fluctuations of $\sigma$ results in fluctuations of its mean value,
\begin{equation}
\bar{N}=\langle N\rangle\frac{\sigma}{\langle\sigma\rangle}.
\label{eq32}
\end{equation}
For fluctuations of cross section given by the gamma distribution~(\ref{eq11}) the resulting multiplicity distribution
\begin{eqnarray}
P\left(N\right) & = & \int_{0}^{\infty}d\bar{N}\frac{\bar{N}^{N}}{N!}\exp\left(-\bar{N}\right) g\left(\bar{N}\right)\nonumber\\
& = & \frac{\Gamma\left(N+k\right)}{\Gamma\left(N+1\right)\Gamma\left(k\right)}\frac{\left(\langle N\rangle/k\right)^{N}}{\left(1+\langle N\rangle/k\right)^{N+k}}
\label{eq33}
\end{eqnarray}
is given by Negative Binomial distribution where $1/k=\omega$. However, in a such approach, the evaluated fluctuations $\omega={\rm Var}\left(\sigma\right)/\langle\sigma\rangle^{2}$ from available data on $1/k=-0.104+0.029\ln s$ are very low for low energy region (in equation~(\ref{eq36}) discussed later, the term $1/\langle N\rangle$ disappear in this case). Alternatively we can assume that multiplicity fluctuations follows the cross section fluctuations, namely
\begin{equation}
N = C\left(s\right) \sigma.
\label{eq34}
\end{equation}
Then
\begin{equation}
{\rm Var}\left(N\right)\cong\left(\frac{dC}{d\sigma}\frac{d\sigma}{ds}\sigma+C\right)^{2}{\rm Var}\left(\sigma\right),
\label{eq35}
\end{equation}
and finally (taking into account the energy dependence of cross section and multiplicity) we can write
\begin{equation}
\omega=\left(\frac{1}{k}+\frac{1}{\langle N\rangle}\right)\xi^{-2},
\label{eq36}
\end{equation}
where $\xi=1+1/\left(2.028-2.67\ln s+0.614\ln^{2} s\right)$.
Comparison of evaluated cross section fluctuations $\omega=Var\left(\sigma\right)/\langle\sigma\rangle^{2}$ from fluctuations of multiplicity is shown in figure~\ref{fig3}.
%

\section{Concluding remarks}
\label{sect:cr}

The fluctuations of cross section is described by the gamma distribution with relative variance $\omega$. This immediately leads to the nucleon-nucleon collision profile given by the regularized incomplete gamma function, which ranges from the hard sphere (for $\omega=0$) to the Gaussian (for $\omega=1$) shape.

Experimental data on elastic differential proton-proton cross section at LHC energy indicate large fluctuations, $\omega=0.4$. Cross section fluctuations are not determined by the spatial distribution of partons but rather by the different configurations of partons. Experimentally observed correlation between mean multiplicity and cross section suggests the possible relation between fluctuations of both observables.

Cross section fluctuations can also affect observables measured in nuclear collisions. It is remarkable that increasing $\omega$ results in decrease of relative fluctuations of participants.

\ack

One of us (MR) wishes to thank Wojciech Broniowski and Przemys\l aw Ko\'{s}cik for useful discussions. This work was supported by the National Science Centre, grant DEC-2011/01/D/ST2/00772.

\end{document}